\documentclass[letter]{ptptex}

\usepackage{graphicx}

\notypesetlogo                       

\title{
Rotating Black Hole 
in Extended Chern-Simons Modified Gravity%
}

\author{
Kohkichi \textsc{Konno},$^{1,2,}$\footnote{E-mail: 
kohkichi@eng.hokudai.ac.jp}
Toyoki \textsc{Matsuyama}$^{3}$
and Satoshi \textsc{Tanda}$^{1,2}$
}
\inst{
$^1$Department of Applied Physics, Hokkaido University,
Sapporo 060-8628, Japan\\
$^2$Center of Education and Research for Topological Science and
Technology, Hokkaido University,
Sapporo 060-8628, Japan\\
$^3$Department of Physics, Nara University of Education,
Nara 630-8528, Japan
}

\recdate{Febrary 27, 2009; Revised April 20, 2009}%

\abst{
 We investigate a slowly rotating black hole  
 in four-dimensional extended Chern-Simons modified gravity.
 We obtain an approximate solution that reduces to the 
 Kerr solution when a coupling constant vanishes.
 The Chern-Simons correction effectively reduces the 
 frame-dragging effect around a black hole 
 in comparison with that of the Kerr solution. 
}

\begin{document}

\maketitle

{\it Introduction}.
There is growing interest in modifications of the theory 
of general relativity. 
The theory of quantum gravity mostly motivates 
such modifications \cite{string,lcg}.
Quantum gravity is requisite for dealing with 
extreme situations in the Universe, such as the vicinity 
of a black hole and the initial state of the Universe. 
For the theory of quantum gravity, several candidates have 
been proposed \cite{string,lcg}. Those candidates certainly 
require modifications of general relativity.
In recent years, serious astrophysical problems of dark matter 
and dark energy also motivate modified gravity theories.
According to recent observational results \cite{wmap}, 
the dark components are considered to comprise over 90\% 
of the energy constituents in the Universe.
Modifications of general relativity would provide 
an alternative way to resolve the unsolved issues
of the dark components.
Among modified gravity theories, we focus on four-dimensional 
Chern-Simons (CS) modified gravity \cite{jp,seck} in this paper.

In CS modified gravity, the action is modified 
by adding a CS term to the Einstein-Hilbert action.
Deser et al. \cite{djt} originally developed the CS
modified gravity in (2+1) dimensions and 
Jackiw and Pi \cite{jp} extended it to (3+1) dimensions 
by introducing an external scalar function.
Furthermore, Smith et al. \cite{seck} have recently considered 
an extended CS modified gravity in which the scalar function 
is regarded as a dynamical scalar field.
The most interesting points of the four-dimensional
CS modified gravity \cite{jp,seck,ad,kmt,kmat,gy,gmm,ys,cs-g} 
are summarized as follows.
First, the CS modified gravity can be obtained 
explicitly from the superstring theory \cite{seck,ad}. 
In the superstring theory, a CS term in the 
Lagrangian density is essential to cancel anomaly. 
Second, as shown by Weinberg \cite{weinberg}, 
the parity-violating higher-order correction 
in an effective field theory for inflation 
also provides the CS modified gravity.
Third, the Schwarzschild solution holds in the 
CS modified gravity \cite{jp}. Thus, the theory 
passes the classical tests of general relativity.
Finally, in the CS modified gravity, the axial part of 
gravitational fields is mainly modified in comparison with 
the cases of general relativity. In fact, the Kerr solution 
does not hold in the CS modified gravity. This is because
the Chern-Pontryagin density, which can be derived from 
the partial integral of the CS term, is closely coupled 
to the axial part of gravitational fields.
In relation to the last point, rotating black hole 
solutions have been discussed in the framework of 
the Jackiw-Pi model \cite{kmt,kmat,gy}.
In Ref.~\citen{kmt}, an approximate solution 
for a rotating black hole was obtained
by using a slow rotation approximation.
The approximate solution interestingly shows the same feature 
as that of flat rotation curves seen in spiral galaxies.\cite{kmat}
In Ref.~\citen{gy}, an exact solution for a rotating 
mathematical black hole was obtained.
Somewhat exotic features of those solutions
are considered to arise from the constraint of vanishing 
Chern-Pontryagin density in the Jackiw-Pi model.
In this paper, we investigate an approximate solution 
for a rotating black hole in the extended CS modified gravity,
in which the constraint is replaced with a field equation 
for a scalar field. To obtain a black hole solution, 
we assume both slow rotation of a black hole and a weak CS 
coupling in this study.
Throughout the paper, we use geometrized units with $c=G=1$.

{\it Extended CS modified gravity}.
The extended CS modified gravity is 
provided by the action \cite{jp,seck} 
\begin{eqnarray}
\label{eq:lagrangian}
 I & = & \int d^4 x \sqrt{-g}
  \left[ - \frac{1}{16\pi}  R + \frac{\ell}{64\pi} 
  \vartheta \: ^{\ast}R^{\tau \ \mu\nu}_{\ \sigma}
  R^{\sigma}_{\ \tau\mu\nu} 
  - \frac{1}{2} g^{\mu\nu} \left( \partial_{\mu} \vartheta \right)
  \left( \partial_{\nu} \vartheta \right) 
  + {\cal L}_{\rm m} \right] , 
\end{eqnarray}
where $g$ is the determinant of the metric $g_{\mu\nu}$, 
$R \equiv  g^{\alpha\beta} R_{\alpha\beta}$ 
($R_{\alpha\beta} \equiv R^{\lambda}_{\ \alpha\lambda\beta}$)
is the Ricci scalar,
$R^{\tau}_{\ \sigma\alpha\beta} \equiv \partial_{\beta}
\Gamma^{\tau}_{\sigma\alpha} - \cdots$ is the Riemann tensor
($\Gamma^{\alpha}_{\beta\gamma}$ is the Christoffel symbols),
$\ell$ is a coupling constant, $\vartheta$ is a dynamical scalar field,
and ${\cal L}_{\rm m}$ is the Lagrangian for matter.
The dual Riemann tensor is defined by
$^{\ast} \!R^{\tau\ \mu\nu}_{\ \sigma} \equiv \frac{1}{2}
\varepsilon^{\mu\nu\alpha\beta} R^{\tau}_{\ \sigma\alpha\beta}$,
where $\varepsilon^{0123} \equiv 1/ \sqrt{-g}$ 
is the  Levi-Civita tensor.
In this paper, we neglect the surface integral term
(see Ref.~\citen{gmm} for a detailed discussion)
and the potential term for $\vartheta$ in the action.
When we neglect the kinematic term of $\vartheta$ 
in Eq.~(\ref{eq:lagrangian}), the action reduces to 
the Jackiw-Pi model developed in Ref.~\citen{jp}.
 From the variations in the action with respect to 
the metric $g_{\mu\nu}$ and the scalar field $\vartheta$, 
we obtain the field equations, respectively, 
\begin{eqnarray}
\label{eq:feq1}
 G^{\mu\nu} + \ell C^{\mu\nu} 
  & = & - 8\pi \left( T_{\rm m}^{\ \ \mu\nu} 
    + T_{\vartheta}^{\ \mu\nu} \right), \\
\label{eq:feq2}
 g^{\mu\nu} \nabla_{\mu} \nabla_{\nu} \vartheta
  & = & - \frac{\ell}{64\pi} \: ^{\ast} R^{\tau \ \mu\nu}_{\ \sigma}
     R^{\sigma}_{\ \tau\mu\nu} ,
\end{eqnarray}
where $G^{\mu\nu}\equiv R^{\mu\nu} - \frac{1}{2} g^{\mu\nu} R$ 
is the Einstein tensor, 
$C^{\mu\nu}$ is the Cotton tensor defined by
\begin{eqnarray}
 C^{\mu\nu} & \equiv & - \frac{1}{2}
 \left[ \left( \nabla_{\sigma} \vartheta \right)
   \left( \varepsilon^{\sigma\mu\alpha\beta}
   \nabla_{\alpha} R^{\nu}_{\ \beta} + 
   \varepsilon^{\sigma\nu\alpha\beta}
   \nabla_{\alpha} R^{\mu}_{\ \beta} \right) +
  \left( \nabla_{\sigma} \nabla_{\tau} \vartheta \right)
   \left( \:\! ^{\ast} \! R^{\tau\mu\sigma\nu} 
    + \:\! ^{\ast} \! R^{\tau\nu\sigma\mu} \right) \right] ,
  \nonumber \\ 
\end{eqnarray}
$T_{\rm m}^{\ \ \mu\nu}$ is 
the energy-momentum tensor for matter, and $T_{\vartheta}^{\ \mu\nu}$ 
is the energy-momentum tensor of the scalar field $\vartheta$, 
\begin{equation}
 T_{\vartheta}^{\ \mu\nu}
 = \left( \nabla^{\mu} \vartheta \right)
   \left( \nabla^{\nu} \vartheta \right)
   - \frac{1}{2} g^{\mu\nu} ( \nabla^{\lambda} \vartheta )
   ( \nabla_{\lambda} \vartheta ) .
\end{equation}
Thus, Eqs.~(\ref{eq:feq1}) and (\ref{eq:feq2}) are basic equations 
for gravitational fields under the extended CS modified gravity.

{\it Approximate solution for a rotating black hole}.
We discuss a slowly rotating black hole by considering
the perturbation of a spherically symmetric static spacetime.
For such a background, we have the metric
$ds^{2} = g^{(0)}_{\mu\nu} dx^{\mu} dx^{\nu}
= - e^{2\Phi (r)} dt^2 + e^{2\Lambda (r)} dr^2 + r^2 
\left( d\theta^2 + \sin^2 \theta d\phi^2 \right)$ 
and the scalar field $\vartheta = \vartheta^{(0)} (r)$,
where $\Phi$ and $\Lambda$ are functions of $r$, 
and the superscript `(0)' denotes the zeroth order in perturbation.
The field equations for the background reduce to \cite{ys}
\begin{eqnarray}
\label{eq:feq1-ss}
 G_{(0)}^{\mu\nu} 
  = - 8\pi T_{\vartheta^{(0)}}^{\ \mu\nu} , \\
\label{eq:feq2-ss}
 g_{(0)}^{\mu\nu} \nabla_{\mu}^{(0)} 
 \nabla_{\nu}^{(0)} \vartheta^{(0)} = 0 .
\end{eqnarray}
Hereafter, we assume $T_{\rm m}^{\ \ \mu\nu} = 0$.
We now adopt the Schwarzschild spacetime with 
$\vartheta^{(0)} = 0$ as a background solution
for Eqs.~(\ref{eq:feq1-ss}) and (\ref{eq:feq2-ss}).
The metric in polar coordinates is given by
\begin{eqnarray}
 g^{(0)}_{\mu\nu} dx^{\mu} dx^{\nu} 
 & = & - \left( 1- \frac{2M}{r} \right) dt^2
 + \left( 1- \frac{2M}{r} \right)^{-1} dr^2 
 + r^2 \left( d\theta^2 + \sin^2\theta d\phi^2 \right) ,
\end{eqnarray}
where $M$ is the mass of a black hole.
Let us consider the perturbation of the Schwarzschild 
spacetime, 
\begin{eqnarray}
 g_{\mu\nu} & = & g^{(0)}_{\mu\nu} + g^{(1)}_{\mu\nu} , \\
 \vartheta & = & \vartheta^{(1)} ,
\end{eqnarray}
where $g^{(1)}_{\mu\nu}$ and $\vartheta^{(1)}$ are at least
first-order quantities of a small parameter $\epsilon$
that characterizes the slow rotation of a black hole, i.e.,
$g^{(1)}_{\mu\nu} , \vartheta^{(1)}
\sim O\left( \epsilon \right)$. 
Thus, we take account of a scalar field that is purely excited by 
the spacetime curvature through Eq.~(\ref{eq:feq2}).
As discussed in detail below, we also adopt the approximation 
of a weak CS coupling, that is, we also regard $\ell$ as 
an expansion parameter. The scalar field obeys Eq.~(\ref{eq:feq2}).
While the right-hand side in Eq.~(\ref{eq:feq2}), i.e.,
$^{\ast} R^{\tau \ \mu\nu}_{\ \sigma}
R^{\sigma}_{\ \tau\mu\nu}$, exactly vanishes for 
spherically symmetric spacetimes, $^{\ast} R^{\tau \ \mu\nu}_{\ \sigma}
R^{\sigma}_{\ \tau\mu\nu}$ becomes nonzero at first order in $\epsilon$.
Thus, the scalar field should be of order $\ell \epsilon $.
Then the CS correction in Eq.~(\ref{eq:feq1}) has order  
$\ell^2 \epsilon$. Therefore, when we take the Schwarzschild 
black hole as a background, the CS correction to the rotation 
of the black hole is of order $\ell^2 \epsilon$.

We discuss the basic equations for such a slowly rotating black hole
in the extended CS modified gravity.
Using the Regge-Wheeler gauge \cite{rw}, we have
the first-order axisymmetric metric as 
\begin{equation}
 g_{\mu\nu}^{(1)} = \left(
  \begin{array}{cccc}
   - \left( 1- \frac{2M}{r} \right) h (r,\theta )
   & H (r,\theta ) &  0
   &  - r^2 \sin^2\theta \omega \left( r , \theta \right) \\
   H (r,\theta ) & \left( 1- \frac{2M}{r} \right)^{-1} m (r,\theta )
   & 0 & - r^2 u \left( r , \theta \right) \\
   0 & 0 & r^2 k \left( r , \theta \right) & 0 \\
   - r^2 \sin^2\theta \omega \left( r , \theta \right)
   & - r^2 u \left( r , \theta \right) & 0
   & r^2 k \left( r , \theta \right) \sin^2\theta
  \end{array}
 \right) ,
\end{equation}
where $\left( h, H, m, k, \omega , u\right) \sim 
O\left( \epsilon\right)$. 
The scalar field is also assumed to depend on $r$ and $\theta$, 
i.e., $\vartheta = \vartheta^{(1)} \left( r,\theta \right) 
\sim O\left( \epsilon \right)$.
We do not pay attention to the order in $\ell$ for a while.
We can obtain differential equations 
for the first-order functions 
$\left( h, H, m, k, \omega , u , \vartheta \right)$
from Eqs.~(\ref{eq:feq1}) and (\ref{eq:feq2}).
Differential equations of first order in $\epsilon$
are completely divided into four groups as 
$\left\{ h, m , k \right\}$, $\{ H \}$, 
$\{ u \}$, and $\left\{ \omega , \vartheta \right\}$.
The $(tt)$, $(rr)$, $(r\theta)$, $(\theta\theta)$,
and $(\phi\phi)$ components of Eq.~(\ref{eq:feq1})
give homogeneous differential equations for 
$h$, $m$, and $k$. The $(tr)$ and $(t\theta)$ components
of Eq.~(\ref{eq:feq1}) 
give homogeneous differential equations for $H$, 
and the $(r\phi)$ and $(\theta\phi)$ components give 
homogeneous differential equations for $u$.
For those components of Eq.~(\ref{eq:feq1}), we have
$C^{\mu\nu} \sim O \left( \epsilon^2 \right)$. 
Of course, a simple solution of $h=m=k=H=u=0$ satisfies 
the differential equations for $h$, $m$, $k$, $H$, and $u$. 
Finally, we obtain differential equations for $\omega$ and $\vartheta$
from the $( t\phi )$ component of Eq.~(\ref{eq:feq1})
and the field equation (\ref{eq:feq2}) for the scalar field,
\begin{eqnarray}
\label{eq:omega}
 \lefteqn{r (r-2M) \partial_{r}^2 \omega 
   + 4 (r-2M) \partial_{r} \omega
   + \partial_{\theta}^2 \omega 
   + 3 \cot\theta \partial_{\theta} \omega} 
   \nonumber \\
 && = \ell \frac{6M (r-2M)}{r^4} \left( \frac{1}{\sin\theta}
   \partial_{r}\partial_{\theta}
   \vartheta - \frac{1}{r\sin\theta} 
   \partial_{\theta} \vartheta \right) , \qquad \\
\label{eq:vartheta}
 \lefteqn{r (r-2M)  \partial_{r}^2 \vartheta 
   + 2 (r-M)  \partial_{r} \vartheta
   + \frac{1}{\sin\theta} \partial_{\theta} \left( 
   \sin\theta \partial_{\theta} \vartheta \right)}
   \nonumber \\
 && = - \ell \frac{3M}{8\pi r} \left( 
   \sin\theta \partial_{r}\partial_{\theta} \omega
   + 2\cos\theta \partial_{r} \omega\right) .
\end{eqnarray}
Equations (\ref{eq:omega}) and (\ref{eq:vartheta})
govern the rotation of a black hole under the 
extended CS modified gravity.

We discuss solutions for Eqs.~(\ref{eq:omega}) and (\ref{eq:vartheta})
to clarify how the black hole rotates.
For this purpose, let us expand $\omega$ and $\vartheta$ as 
\begin{eqnarray}
 \omega (r,\theta ) 
  & = & \sum_{n=1}^{\infty} \tilde{\omega}_{n} (r) 
        \frac{1}{\sin\theta} \partial_{\theta} 
        P_{n} (\cos\theta ) , \\
 \vartheta (r,\theta ) 
  & = & \sum_{n=1}^{\infty} \tilde{\vartheta}_{n} (r) 
        P_{n} (\cos\theta ) ,
\end{eqnarray}
where $P_{n}$ denotes the Legendre function of the first kind.
 From Eqs.~(\ref{eq:omega}) and (\ref{eq:vartheta}), 
we obtain for each $n$
\begin{equation}
\label{eq:omega-r}
 \tilde{\omega}_{n}'' + \frac{4}{r} \tilde{\omega}_{n}'
  + \frac{2-n(n+1)}{r(r-2M)} \tilde{\omega}_{n}
 = \ell \frac{6M}{r^5} \left( \tilde{\vartheta}_{n}'
    - \frac{\tilde{\vartheta}_{n}}{r} \right) , 
\end{equation}
\begin{equation}
\label{eq:vartheta-r}
 \tilde{\vartheta}_{n}'' + \frac{2(r-M)}{r(r-2M)} 
 \tilde{\vartheta}_{n}'
  - \frac{n(n+1)}{r(r-2M)} \tilde{\vartheta}_{n}
 = \ell \frac{3n(n+1)M}{8\pi r^2 (r-2M)} 
     \tilde{\omega}_{n}' ,
\end{equation}
where the prime denotes the differentiation with respect to $r$.
To solve Eqs.~(\ref{eq:omega-r}) and (\ref{eq:vartheta-r}), 
let us assume $\ell \ll 1$. 
We solve Eqs.~(\ref{eq:omega-r}) and (\ref{eq:vartheta-r})
by the method of iteration by considering $\ell$ 
to be an expansion parameter.
The homogeneous differential equations of 
Eqs.~(\ref{eq:omega-r}) and (\ref{eq:vartheta-r})
give the equations of order $\ell^{0} \epsilon$.
We obtain the homogeneous solutions as
\begin{eqnarray}
\label{eq:omega_n}
 \tilde{\omega}_{n} 
 & = & 
   A_{0} \: _{2}F_{1} \left( 1-n , 2+n , 4 ; \frac{r}{2M} \right) 
   + B_{0} \: G^{2,0}_{2,2} \left( \left. \frac{r}{2M} \right|
   \begin{array}{ccc}
    -n-1 & , & n \\
    -3 & , & 0
   \end{array}
   \right) , \\
\label{eq:vartheta_n}
 \tilde{\vartheta}_{n} 
 & = & C_{0} \: P_{n} \left( \frac{r}{M} -1 \right)
     + D_{0} \: Q_{n} \left( \frac{r}{M} -1 \right) , 
\end{eqnarray}
where $A_{0}$, $B_{0}$, $C_{0}$, and $D_{0}$ are constants, 
$_{2}F_{1}$ is a hypergeometric function, 
$G^{2,0}_{2,2}$ is a Meijer G-function \cite{meijer}, and
$Q_{n}$ is the Legendre function of the second kind.
Note that while $P_{n} \left( r/M -1 \right)$ diverges
at $r\rightarrow \infty$, 
$Q_{n} \left( r/M -1 \right)$ diverges at $r=2M$.
Thus, we should have $C_{0}=D_{0}=0$ at order $\ell^{0} \epsilon$.
Otherwise, $\tilde{\omega}_{n}$ would also diverge at $r=2M$ or
$r\rightarrow \infty$ at higher orders.
The metric solution in Eq.~(\ref{eq:omega_n}) for $n=1$
gives the linear approximation of the Kerr solution.
The solution for $n=1$ allows us to compare a rotating black hole
in the extended CS modified gravity with that in general relativity.
Hence, we focus on the solution for $n=1$ hereafter.
The solution of order $\ell^{0} \epsilon$ is summarized as
\begin{equation}
\label{eq:zeroth}
 \tilde{\omega}_{1} = - \frac{2J}{r^3} , 
 \quad \tilde{\vartheta}_{1} = 0 , 
\end{equation}
where $J$ is the angular momentum of a black hole.
Substituting Eq.~(\ref{eq:zeroth}) into the right-hand side
of Eq.~(\ref{eq:vartheta-r}), 
we obtain a solution of order $\ell \epsilon$,
\begin{eqnarray}
\label{eq:vartheta-1st}
 \tilde{\vartheta}_{1} 
 & = & C_{1} \: P_{1} \left( \frac{r}{M} -1 \right)
     + D_{1} \: Q_{1} \left( \frac{r}{M} -1 \right) \nonumber \\
 && - \ell \frac{J}{256 \pi M^5 r^4} \biggl[  2M 
  \left( 15r^4 + 5M^2 r^2 + 10M^3 r + 18M^4 \right)  
  \nonumber \\
 && \left. 
  - 15 r^4 (r-M) \ln \left| \frac{r}{r-2M} \right|\right] ,
\end{eqnarray}
where $C_{1}$ and $D_{1}$ are constants.
When we impose the regularity of $\tilde{\vartheta}_{1}$
both at $r=2M$ and at $r\rightarrow \infty$, we derive
\begin{equation}
 C_{1} = 0 , \quad D_{1} = - \ell \frac{15J}{256\pi M^5} .
\end{equation}
Thus, we obtain the regular solution of order $\ell \epsilon$ as
\begin{equation}
\label{eq:vartheta-1st-f}
 \tilde{\vartheta}_{1}
 = - \ell \frac{J}{128\pi M^2 r^4} 
   \left( 5r^2 + 10Mr + 18M^2 \right) .
\end{equation}
In a similar way, substituting this solution into 
the right-hand side of Eq.~(\ref{eq:omega-r}), 
we obtain the solution of order $\ell^2 \epsilon$ as
\begin{equation}
\label{eq:omega-2nd}
 \tilde{\omega}_{1}
 = \ell^2 \frac{J}{1792\pi M r^8}
   \left( 70r^2 + 120Mr +189M^2 \right) .
\end{equation}
Consequently, we obtain the approximate solution up to 
order $\ell^2 \epsilon$ for a slowly rotating black hole,
\begin{eqnarray}
\label{eq:omega-final}
 \omega
 & = & \frac{2J}{r^3} 
   \left[ 1 - \ell^2 \frac{1}{3584\pi M r^5}
   \left( 70r^2 + 120Mr 
   +189M^2 \right) \right] , \\
\label{eq:vartheta-final}
 \vartheta
 & = & - \ell \frac{J}{128\pi M^2 r^4} 
   \left( 5r^2 + 10Mr + 18M^2 \right) \cos\theta .
\end{eqnarray}
 From Eq.~(\ref{eq:omega-final}), we obtain the metric 
solution for a slowly rotating black hole under 
the extended CS modified gravity
\begin{eqnarray}
\label{eq:metric-final}
 ds^{2}
 & = & - \left( 1- \frac{2M}{r} \right) dt^2 
 + \left( 1- \frac{2M}{r} \right)^{-1} dr^2 
 + r^{2} d\theta^2 + r^2 \sin^2 \theta d\phi^2 
   \nonumber \\
 && - \frac{4J}{r} 
   \left[ 1 - \ell^2 \frac{1}{3584\pi M r^5}
   \left( 70r^2 + 120Mr +189M^2 \right) \right] 
   \sin^2 \theta dtd\phi .
\end{eqnarray}
At infinity, the $(t\phi)$ component of the metric and  
scalar field reduce to 
\begin{eqnarray}
\label{eq:g_tphi_infty}
 g_{t\phi}
 & \simeq & - \frac{2J}{r} 
   \left( 1 - \ell^2 \frac{5}{256\pi M r^3} \right)
   \sin^2 \theta , \\
\label{eq:vartheta_infty}
 \vartheta
 & \simeq & - \ell \frac{5J}{128\pi M^2 r^2} \cos\theta .
\end{eqnarray}
At $r\simeq 2M$, we have
\begin{eqnarray}
\label{eq:g_tphi_2M}
 g_{t\phi}
 & \simeq & - \left[ \frac{J}{M} 
   \left( 1 - \ell^2 \frac{709}{114688\pi M^4} \right) 
   - \frac{J}{2M^2} \left( 1- \ell^2 
   \frac{1727}{57344\pi M^4} \right) (r-2M)\right]
   \sin^2 \theta , \nonumber \\ \\
\label{eq:vartheta_2M}
 \vartheta
 & \simeq & - \ell J \left[ \frac{29}{1024\pi M^4} 
  - \frac{43 (r-2M)}{1024\pi M^5} \right] \cos\theta .
\end{eqnarray}
As shown in Eqs.~(\ref{eq:g_tphi_infty})--(\ref{eq:vartheta_2M}), 
both the metric and scalar field are regular 
at $r\rightarrow \infty$ and at $r=2M$. 
 From Eqs.~(\ref{eq:metric-final}), (\ref{eq:g_tphi_infty}),
and (\ref{eq:g_tphi_2M}), we also find that 
the higher-order correction effectively reduces 
the frame-dragging effect around the black hole
compared with the first-order approximation of 
the Kerr black hole. This result provides contrast to the result 
obtained in Refs.~\citen{kmt} and \citen{kmat} within the Jackiw-Pi model. 
The reduction of the frame-dragging effect 
is attributed to the existence of 
the kinematic term of the scalar field.

\begin{figure}[tb]
 \begin{center}
  \includegraphics[width=0.9\textwidth]{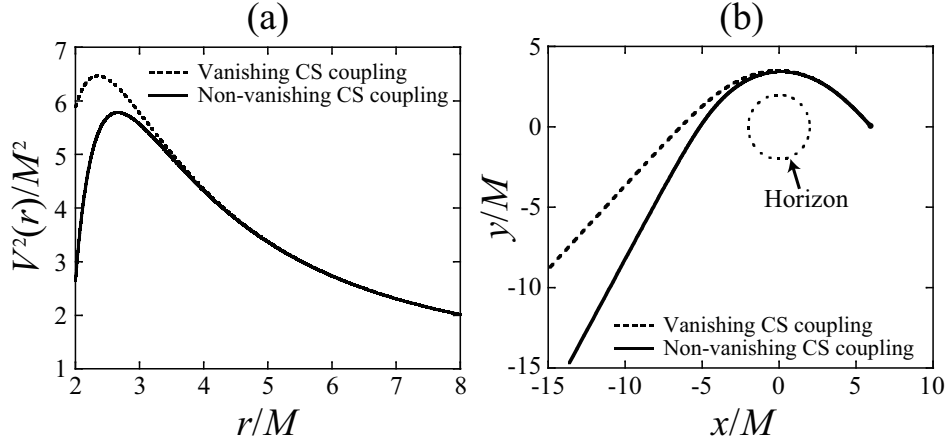}
 \end{center}
 \caption{ 
  (a) Potential $V^{2}(r)$. 
  (b) Particle trajectories in the equatorial plane. 
  The starting point is indicated by a small filled circle.
  The results in the presence of the CS coupling 
  are denoted by solid curves, while the results in the 
  absence of the CS coupling are denoted by dashed curves.
  In both figures, we assume $E=2.35$, $L/M=10.0$, 
  $J/M^2 = 0.5$, and $\tilde{\ell}^2=0.1$.}
 \label{fig1}
\end{figure}

{\it Observational implications}.
We discuss observational implications for the 
black hole solution given in Eq.~(\ref{eq:metric-final}). 
The CS correction in $g_{t\phi}$ has the same polar angle 
dependence as that of the first-order rotational approximation of 
the Kerr solution. Thus, the CS correction can be found 
only from the radial dependence of $g_{t\phi}$.
The CS correction is composed of higher-order terms of $1/r$,
i.e., at least terms of $O\left( 1/r^4 \right)$.
This fact denotes that the CS correction effectively modifies
the frame-dragging effect in the vicinity of the black hole.
Therefore, to provide an upper limit for the CS coupling, it is 
useful to observe an object passing close by the black hole.
Let us consider a massive particle with mass $m$ and
four-momentum $p^{\mu}$ on the equatorial plane of
the black hole spacetime. 
For a particle with energy $E=-p_{t}/m$ and angular momentum 
$L=p_{\phi}/m$, from $p^{\mu}p_{\mu} = -m^2$, we obtain
\begin{equation}
\label{eq:dr_dtau}
 \left(\frac{dr}{d\tau} \right)^2
 = E^2 - V^{2}(r) ,
\end{equation}
where $\tau$ is the proper time of the particle, and 
\begin{eqnarray}
 V^{2} & = & \left( 1- \frac{2M}{r} \right)
  \left( 1+\frac{L^2}{r^2} \right) + LEJ
  \left[ \frac{4}{r^3} - \tilde{\ell}^2 \frac{1}{M r^8}
   \left( 70r^2 + 120Mr +  189M^2 \right)\right] .
   \nonumber \\
\end{eqnarray}
Here, we define ${\tilde{\ell}}^{2} \equiv \ell^2/896\pi$.
Particle trajectories are permitted for a region 
where $E^2 > V^{2}$. Figure \ref{fig1}(a) shows the potential
under a certain condition with the CS coupling (solid curve)
and that without the CS coupling (dashed curve). 
In this figure, we assume corotating trajectories.
The CS correction reduces the potential 
for a region close to the horizon $r=2M$.
At the peaks of the potentials, unstable circular orbits 
are realized if $E^2 = V^2 (r_{\rm peak})$. 
As seen in Fig.~\ref{fig1}(a), the peak is shifted outward 
by the CS correction. However, when a particle that 
has energy of $E^{2}< V^2 (r_{\rm peak})$ falls freely  
from the outside ($r>r_{\rm peak}$), the periastron is shifted 
inward by the CS reduction of the potential.
Multiplying Eq.~(\ref{eq:dr_dtau}) by $(d\phi/d\tau)^{-2}$ and 
differentiating that with respect to $\phi$, we obtain
the differential equation for trajectories
\begin{eqnarray}
\label{eq:diff_traj}
 \frac{d^2r}{d\phi^2}
 & = & \frac{1}{L^2} 
   \left( 2E^2 r^3 -2r^3 +3Mr^2 -L^2 r + ML^2 \right) 
   +\frac{2J}{L^3} \left[ 2Er^2 
   - E^3 \frac{(3r-8M)r^3}{(r-2M)^2} \right]
    \nonumber \\
 & & +\tilde{\ell}^2 \frac{J}{L^3} 
   \left[ E \frac{3(20r+63M)}{r^3} - E^3 
   \frac{130r^2+189Mr-189M^2}{r^2(r-2M)^2} \right] .
\end{eqnarray}
We can obtain particle trajectories by solving this equation.  
Figure \ref{fig1}(b) shows trajectories
for the potentials given in Fig.~\ref{fig1}(a).
We used the Runge-Kutta method to integrate 
Eq.~(\ref{eq:diff_traj}) numerically.
As expected, we find that the periastron 
is closer to the black hole
in the presence of the CS coupling than its absence.
As seen from Fig.~\ref{fig1}, the CS correction effectively 
modifies the trajectory around the periastron. 
The effect would also be magnified at the apastron.
Therefore, observations of a satellite in an extended elliptical 
orbit, for example, a system such as OJ287,\cite{oj287}
would provide useful information about the CS coupling.

{\it Concluding remarks}.
We have investigated a rotating black hole
in the extended Chern-Simons (CS) modified gravity.
We considered the perturbation of the Schwarzschild 
spacetime with a vanishing scalar field to discuss 
the rotation of a black hole. 
The assumption of the vanishing scalar field 
at the zeroth order allows us to compare 
a result with that of the Kerr solution.
By taking account of the scalar field that is purely 
excited by a gravitational field through the 
Chern-Pontryagin density, we obtained an approximate 
solution for a slowly rotating black hole.
In obtaining the solution, we adopt both the slow rotation 
approximation and the approximation of a weak CS coupling.  
Consequently, we found that the CS correction gives
an effective reduction of the frame-dragging around
a black hole in comparison with that of the Kerr solution.
We also showed that observations of systems that have 
a satellite in an extended elliptical orbit would be useful 
to provide an upper limit for the CS coupling. 
As a future work, it would be interesting to seek exact 
solutions for a rotating black hole in the extended 
CS modified gravity. From such investigations,
the nonlinear characteristics of the CS modified gravity
would also be revealed.

\section*{Acknowledgements}
One author (K.K.) thanks Prof. R. Jackiw for his hospitality 
at Massachusetts Institute of Technology. 
This work was supported in part by a Grant-in-Aid
for Scientific Research from the 21st Century COE
Program ``Topological Science and Technology''.
Analytical calculations were performed in part 
on computers at YITP of Kyoto University. 

%

Note added. The same solution for the metric was also derived
independently in Ref.~\citen{yp}.

\end{document}